\documentclass[epj,final,floatfix]{svjour}
\usepackage{graphicx}

\begin{document}
\title{Associative memory on a small-world neural network}

\date{\today}

%\author{Luis G. Morelli}
%\email{morelli@ictp.trieste.it}
%\affiliation{Abdus Salam International Center for Theoretical Physics, P.O. Box 586, 34100 Trieste, Italy}
%\author{Guillermo Abramson}
%\email{abramson@cab.cnea.gov.ar}
%\affiliation{Centro At\'{o}mico Bariloche, CONICET and Instituto Balseiro, 8400 S. C. de Bariloche, Argentina}
%\author{Marcelo N. Kuperman}
%\email{kuperman@cab.cnea.gov.ar}
%\affiliation{Centro At\'{o}mico Bariloche, CONICET and Instituto Balseiro, 8400 S. C. de Bariloche, Argentina}

\author
{
Luis G. Morelli\inst{1,}\thanks{E-mail address: morelli@ictp.trieste.it}
\and Guillermo Abramson\inst{2,}\thanks{E-mail address: abramson@cab.cnea.gov.ar}
\and Marcelo N. Kuperman\inst{2,}\thanks{E-mail address: kuperman@cab.cnea.gov.ar}
}

\institute
{Abdus Salam International Center for Theoretical Physics,
P.O. Box 586, 34100 Trieste, Italy
\and
Centro At\'{o}mico Bariloche, CONICET and Instituto Balseiro,
8400 S. C. de Bariloche, Argentina
}

\abstract{We study a model of associative memory based on a neural
network with small-world structure. The efficacy of the network to
retrieve one of the stored patterns exhibits a phase transition at
a finite value of the disorder. The more ordered networks are unable to
recover the patterns, and are always attracted to mixture states.
Besides, for a range of the number
of stored patterns, the efficacy has a maximum at an intermediate
value of the disorder. We also give a statistical characterization
of the attractors for all values of the disorder of the network.
\PACS { {84.35.+i}{Neural networks}\and
{89.75.Hc}{Networks and genealogical trees}\and {87.18.Sn}{Neural
networks} } }

\authorrunning{L. G. Morelli, G. Abramson and M. N. Kuperman}

\maketitle

\section{Small-world neural networks}

Artificial neural networks have been used as a model for
associative memory since the 80's, and a considerable a\-mount of work
has been made in the field \cite{amit,peretto}. Most of this work
regards both the simulation and the theory of completely connected
networks, as well as networks with a random dilution of the
connectivity. It is known that particular prescriptions for the
determination of the synaptic weights enable these systems to
successfully retrieve a pattern out of a set of memorized ones.
This behavior is observed in the system up to a certain value of
the number of stored patterns, beyond which the network becomes
unable to retrieve any of them. For reasons of simplicity of the
models and their analytical tractability, complex architectures of
the networks, more akin to those found in biological neural
systems, have been largely left out of the theoretical analysis.
Fortunately, since a few years ago, a class of models that has
come to be known as ``complex networks'' began to be thoroughly
studied. Complex networks seem more compatible with the
geometrical properties of many biological and social phenomena
than regular lattices, random networks, or completely connected
systems \cite{watts1998,barabasi99,newman00,watts}.
Already in the seminal work of Watts and
Strogatz \cite{watts1998}, whose small-world model combines
properties of regular and random networks, it was observed
that the neural system of the nematode \textit{C. elegans} shares
topological properties with this model networks.

In this paper we study a neural network built upon the
Watts-Strogatz model for small worlds. The model interpolates
between regular and random networks by means of a parameter $p$,
which characterizes the disorder of the network. The construction,
as formulated in Ref. \cite{watts1998}, begins with a
one-dimensional regular lattice of $N$ nodes, each one linked to
its $K$ nearest neighbors to the right and to the left, and with
periodic boundary conditions. With probability $p$, each one of
the right-pointing links, of every node, is rewired to a randomly
chosen node in the network. Self connections and repeated
connections are not allowed. The result is a disordered network,
defined by the set $N$, $K$, $p$, that lies between a regular
lattice ($p=0$) and a random graph ($p=1$). A wide range of these
networks displays high local clusterization and short average
distance between nodes, as many real complex networks. They can be
defined by the \textit{connectivity matrix} $c_{ij}$, where
$c_{ij} = 1$ if there is a link between nodes $i$ and $j$, and
$c_{ij} = 0$ otherwise.  
We use this matrix to establish the synaptic
connections between neurons, at variance from the traditional
Hopfield model, where the network is completely connected and the
connectivity matrix is $c_{ij}=1$, $\forall i,j$. 
At $p=1$ it coincides with
the standard diluted disordered networks, that have also
been considered in the literature, in which randomly
chosen elements in the connectivity matrix are set to zero.

The biological neuron carries out an operation on the inputs
provided by other neurons, and it produces an output. A
transformation of this continuous output into a binary variable
makes it possible to formulate a simplified model in which the
neurons are logical elements (Ref. \cite{amit}, chapter 2).
In this binary representation, the
state of each neuron is characterized by a single variable $s_i$.
This variable can take two values representing the active and the
inactive states,
\begin{equation}
\label{eqstate}
s_i = \left\{
\begin{array} {rl}
1  & \, \mbox{ if the neuron is active}, \\
-1 & \, \mbox{ if the neuron is inactive}.
\end{array}
 \right.
\end{equation}

The purpose of an associative memory model, is to retrieve some patterns
that have been stored in the network by an unspecified learning process.
The stored---or \textit{memorized}---patterns are represented by network states
$\xi^{\mu}$, where $\mu = 1,\dots,M$ labels the different
patterns and $M$ is their number. As usual, the patterns are
generated at random, assigning with equal probability $1/2$ the
values $\xi^{\mu}_i = \pm 1$. The patterns are uncorrelated and
thus orthogonal in large networks:
\begin{equation}
\frac{1}{N} \sum_{i=1}^{N} \xi^{\mu}_i \xi^{\nu}_i =
\delta_{\mu\nu}.
\end{equation}

The state of the neurons is updated asynchronously, as in Glauber
dynamics. At each simulation step, a neuron is chosen at random,
and its new state is determined by the local field:
\begin{equation} \label{eqhi}
h_i = \sum_{j=1}^{N}{\omega_{ij} \, s_j},
\end{equation}
according to:
\begin{equation} \label{eqsi}
s_i = \mbox{sign} \, ({h_i}).
\end{equation}

The synaptic weights $\omega_{ij}$ of the connections are given by
Hebb's rule, restricted to the synapsis actually present in the
network, as given by the connectivity matrix:
\begin{equation} \label{eqwij}
w_{ij} = \frac{1}{N} \sum_{\mu = 1}^{M}{c_{ij}
\xi^{\mu}_{i}\xi^{\mu}_{j}}
\end{equation}
for $i,j=1,\dots,N$. Note that as the network model does not allow
self connections the diagonal matrix elements are null. By
definition, the synaptic matrix is symmetric.

The dynamics prescribed by Eqs.~(\ref{eqhi}) and~(\ref{eqsi}) is
deterministic, and the network is not subject to thermal
fluctuations. We will only consider the effects of a small amount
of additive noise to verify the robustness of our results. A full
discussion of the effect of a finite temperature in the dynamics
will be left for future work. The stochastic asynchronous update,
though, prevents the system from having limit cycles, and the
only attractors are fixed points. The stored patterns $\xi^{\mu}$
are, by construction of the synaptic weights~(\ref{eqwij}), fixed
points of the dynamics due to the orthogonality condition. In the
model, ``memory''  is the capacity of the network to retrieve one
of the stored patterns from an arbitrary initial condition. As
in traditional models, the reversed patterns $(-\xi_i)$, as well
as a wealth of symmetric and asymmetric mixtures of patterns, are
also equilibria of the system and play a significant role in its
behavior as a memory device.

\section{Effect of the disordered topology}

\begin{figure}
\centering \resizebox{\columnwidth}{!}{\rotatebox{-90}
{\includegraphics{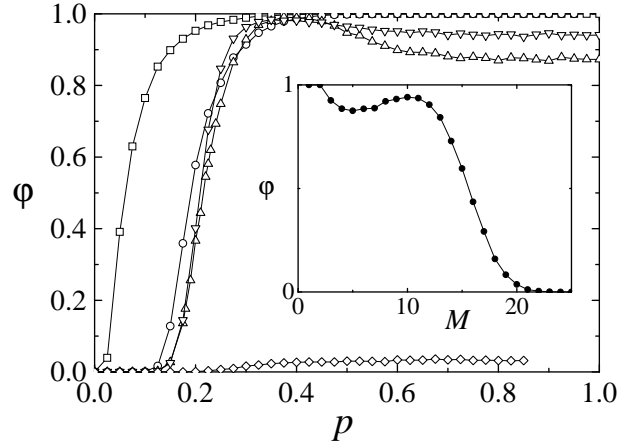}}} \caption{Efficacy to retrieve a
memorized pattern, $\varphi$, as a function of the disorder $p$.
The curves correspond to different number of stored patterns:
(squares) $M=1$, (circles) $M=2$, (up triangles) $M=5$, (down
triangles) $M=10$,  (diamonds) $M=20$. Inset: The efficacy as a
function of the number of stored patterns, at $p=1$. Simulation
parameters: $N=5 \times 10^3$, $K=100$, $10^4$ realizations per point.}
\label{npat}
\end{figure}

We have performed extensive numerical simulations of the system,
starting from a random unbiased initial condition. After a
transient, a fixed point is reached, whence no further changes
occur to any neuron. In order to measure the efficacy of the
network to recall a number $M$ of stored random patterns, we
define an \textit{efficacy} $\varphi$ as the fraction of
realizations in which one of the stored patterns is retrieved. In
Fig.~\ref{npat} we plot the order parameter $\varphi$ as a
function of the disorder parameter $p$.  The different curves
correspond to different numbers of stored patterns, $M=1$, $2$,
$5$, $10$, and $20$. For this plot we have used $N=5 \times 10^3$ and
$K=100$. Averages have been taken over $10^4$ realizations. For
each realization we use different patterns, as well as different
initial conditions. Figure~\ref{npat} shows that on highly
ordered networks the system does not retrieve any stored pattern.
Then there is a transition as the disorder parameter $p$ grows,
and above some critical value of $p$, patterns are retrieved as
fixed points yielding $\varphi >0$. For $M=1$ and $M=2$, $\varphi
\equiv 1$ above $p\approx 0.4$. But for $M>2$ we find that
$\varphi$ does not grow monotonically with $p$. Instead, it
decays as $p$ grows after reaching a maximum value. This
surprising non monotonic behavior with the disorder parameter $p$
has been observed before in a problem of biased
diffusion~\cite{zanette02}, and in an Ising
model~\cite{sanchez2002}, both with asymmetric interactions.

In the inset of Fig.~\ref{npat} we plot $\varphi$ vs. $M$ for a
disordered network with $p=1$. As the number of stored patterns
$M$ grows, the network is not able to retrieve them. The curve
also shows a non monotonic behavior with $M$. The transition as
the number of stored patterns grows has already been studied in
diluted disordered networks (Ref. \cite{amit}, chapter 7).  It is
known that random dilution reduces capacity of a neural network
in a way which is proportional to the fraction of available
connections. For our system (which is very diluted) the
transition, then, takes place at $M_c\approx 0.15 (K/N) N= 15$,
as observed. Nevertheless, we are mostly interested in the
behavior of the system regarding the different topologies
characterized by $p$. The fact that the transition between the
remembering and the non-remembering phases occurs at a finite
value of the disorder parameter is very interesting, since a few
dynamical systems based on small-world architectures show it
\cite{kuperman2001,zanette2002,szabo2003}. This occurs in spite
of the fact that the average distance between nodes, the main
geometrical property of the Watts-Strogatz model, has a
transition at $p=0$ \cite{barrat}. Indeed, for several Ising-like
systems, which bear some similarities with artificial neural
networks, a phase transition occurs at $p=0$~\cite{barrat,zhu03,herrero02,kim01}.

In order to understand the finite size effects in the system, and
the behavior of the transition in the limit of an infinite system,
we have made simulations on systems of different sizes. We have
chosen to keep the connectivity parameter of the model constant
through all the results we show, $K=100$. In this regard, our
results correspond to a neural network characterized by certain
properties at the local level, for example the average
connectivity of each neuron ($2K$ in our systems). Our finite size
analysis shows the behavior of these networks in systems of
increasing size $N$ and in the limit $N\to\infty$.

\begin{figure}
\centering \resizebox{\columnwidth}{!}{\rotatebox{-90}
{\includegraphics{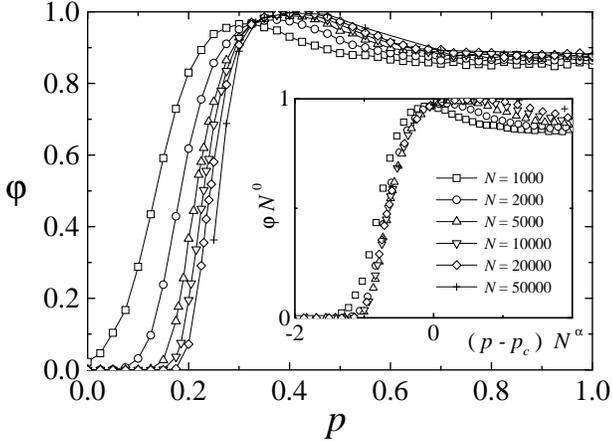}}} \caption{Efficacy $\varphi$ as a
function of the disorder parameter $p$, for systems of different
sizes (as shown in the legend), and $K=100$. The number of stored
patterns is $M=5$, with $10^4$ realizations per point. Inset: The
same curves, scaled with the system size according to
Eq.~(\ref{eqscaling}), collapse to a single curve $\Phi$, with
$p_c=0.333$ and $\alpha=0.2$.} \label{fidepe}
\end{figure}

The plot of $\varphi$ vs. $p$ for different values of $N$ is shown
in Fig.~\ref{fidepe}. For this curves we have set $K=100$ and
$M=5$, averaging over $10^4$ independent realizations. As seen in
the figure, all the curves seem to cross for the same value of the
disorder parameter $p=p_c\approx 0.333$.

Based on numerical evidence, we find that the dependence of the
efficacy on the system size can be built into a scaling function:
\begin{equation}
\varphi \left( p,N \right) = \Phi \left[ (p-p_c) N^{\alpha}
\right].
\label{eqscaling}
\end{equation}
At the point of crossing of the curves,  $\varphi$ becomes
independent of $N$.

Since the order parameter is not singular at the transition,
we can expand $\Phi$ as a Taylor series around the critical
control parameter $p_c$:
\begin{equation}
\varphi \left( p,N \right) = \Phi (0) + \Phi' (0) \, (p-p_c) \,
N^{\alpha},
\end{equation}
to first order in $(p-p_c)$. Defining $\tilde \varphi = \varphi -
\varphi (p_c)$ and $\tilde p = p - p_c$ we can write:
\begin{equation}
\left. \frac{\partial \tilde \varphi}{\partial \tilde p}(N)
\right|_{\tilde p = 0} = \Phi' (0) \, N^{\alpha}.
\end{equation}

Plotting on a log-log scale the derivative ${\partial \tilde
\varphi} / {\partial \tilde p} |_0$ vs. $N$, we obtain the exponent
$\alpha$ as the slope of the line. 
Using data from $N=2 \times 10^3$ to $N=10^5$,
we find $\alpha = 0.23 \pm 0.04$, and $\Phi'(0) = 0.096 \pm 0.016$.
In the inset of Fig.~\ref{fidepe}, we plot the re-scaled curves for different $N$.
The best data collapse is obtained with $\alpha = 0.2$, compatible with the above result.
Observe that the data corresponding to $N=10^3$ (squares) fail to match the scaling curve,
indicating a lower bound of what can be considered a ``large'' system for this model.

Except in the relatively narrow range of $p$ where $\varphi \approx
1$, the system fails to retrieve any stored pattern in a
significant fraction of the realizations: almost always when the
network is very ordered (down to $p=0$), and about 12\% of the
times when the network is very disordered (up to $p=1$). What
happens in the phase space as the network architecture changes?
What happens to the trajectories, and why are the patterns
missed? It seems natural to expect that the energy landscape is
different for $p=0$ than for $p=1$. To address this problem we
turn our attention to the properties of the overlaps of the
equilibrium state with the memorized patterns. Suppose that after
a transient the network has reached a fixed point $\zeta$. We
define the overlap of this fixed point with the patterns as
\begin{equation}
\theta^{\mu} = \frac{1}{N} \left|\sum_{i = 1}^{N}{\xi^{\mu}_i
\zeta_i}\right|.
\end{equation}

\begin{figure}[b]
\centering
\resizebox{\columnwidth}{!}{\rotatebox{-90}{\includegraphics{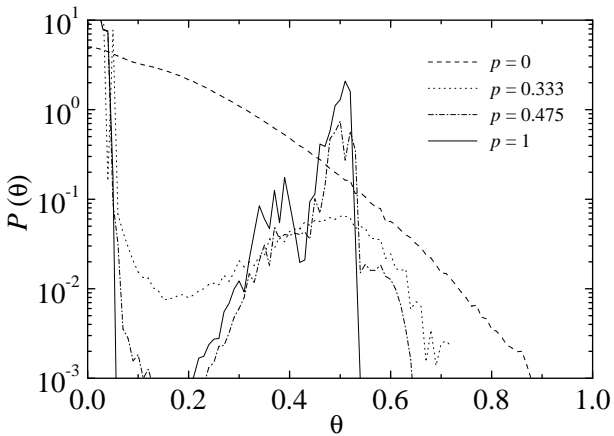}}}
\caption{Distribution of overlaps $P(\theta)$ after a fixed point has been
achieved, between the state of the system and all stored patterns. Each curve
corresponds to a value of $p$, as shown in the legend, typical of the different
memory behaviors observed. A large peak at $\theta=1$ (perfect retrieval of a
pattern) is not shown for reasons of scale (see discussion in the text).}
\label{ovldis}
\end{figure}

Note that if the fixed point is a stored pattern, $\zeta =
\xi^{\nu}$, then $\theta^{\nu}=1$. In order to determine the type
of fixed points that are reached when the network misses the
patterns, we measure the overlap $\theta ^{\mu}$ of the fixed
point with the stored patterns $\xi^{\mu}$. The probability
distribution $P(\theta)$ of these overlaps gives information on
the kind of mixture that the fixed point is. Figure~\ref{ovldis}
shows the overlap distributions for several levels of disorder in
the network. In this plots, $N=2 \times 10^3$, $K=100$, $M=5$ and $10^6$
realizations are used per curve. For the three higher values of
$p$, the distributions have a high peak at $\theta = 1$, which is
not shown for reasons of scale. This peak corresponds to the
realizations that end up in a pattern, which happens frequently
whenever $p>p_c$, as seen in Fig.~\ref{fidepe}. The somewhat
broader peak that these distributions have at low values of
$\theta$ has the same origin, since the overlaps with the other
$M-1$ patterns have a low value whenever a pattern is reached.
Indeed, the overlap of two uncorrelated states has a mean value
$\theta_0=0.022$. In the intermediate range of $\theta$, the
distribution presents a broad bump around $\theta = 1/2$. This
corresponds to symmetric mixtures of the patterns, although the
width of this bump suggests that asymmetric mixtures are present
as well. In particular, the smaller peak present around
$\theta\approx 0.35$ for the completely random network,
corresponds to asymmetric mixtures. In contrast with these three
cases---at and above the critical point---for ordered networks
with $p=0$ the overlap distribution is broad and does not have
peak at $\theta =1$. It has a maximum at $\theta = 0$ and decays
as $\theta$ grows, but large overlaps are observed in some
realizations as the distribution shows. This is the only curve for
which the complete distribution is shown. As the distribution
suggests, the fixed points of these systems consist of very
asymmetric mixtures.

The previous analysis unveiled the structure of the phase space
and the difference between the low and the high $p$ regimes.
Still, what is the reason for the catastrophic loss of memory
below the critical value of disorder? We have found that, for low
values of disorder, the fixed points retrieve scattered pieces of
several stored patterns. These fixed points consist of localized
regions that overlap with different patterns.  Indeed, at $p=0$,
the network is topologically very clusterized, and there exist
local neighborhoods relatively isolated from each other. These
neighborhoods begin to disappear by the action of the shortcuts
provided by the random rewiring at higher values of $p$, until the
whole system becomes essentially a single neighborhood. Then, at
$p=0$, from an arbitrary initial condition, different regions of
the network eventually align themselves with different patterns.
The final result is a completely asymmetric mixture, impossible to
classify due to the arbitrariness of its origin and nature. These
are the states that the broad distribution of overlaps describes,
in Fig.~\ref{ovldis}, for $p=0$. The existence of asymmetric
mixtures as attractors in this kind of associative memory model
have been observed before (see for example \cite{amit}, chapter
4). But since they are very rare in the completely random or in
the completely connected networks, they are very difficult to
observe. In the present context, however, they play an essential
role in the destruction of the ability of the system to retrieve
the patterns.

In order to quantify this, we proceed to define a correlation
measure that provides a clear picture of the situation. We
introduce the difference of the fixed point $\zeta$ with a given
pattern:
\begin{equation}
d^{\mu}_i =  \xi^{\mu}_i \zeta_i =
\left\{
\begin{array} {rl}
1  & \, \mbox{ if } \quad \xi^{\mu}_i =    \,\zeta_i,\\
-1 & \, \mbox{ if } \quad \xi^{\mu}_i \neq \,\zeta_i.
\end{array}
 \right.
\end{equation}
Then we define a local magnetization for the difference vector
$d^{\mu}$, for every node $i$:
\begin{equation}
m^{\mu}_i = \frac{1}{1+k_i} \left| d^{\mu}_i + \sum_{j \in {\mathcal V}_i}
{d^{\mu}_j} \right|,
\end{equation}
where ${\mathcal V}_i$ is the set of neighbors of node $i$. The
local magnetization $m^{\mu}_i$ measures the local alignment with
the $\mu$ pattern or its reversed companion. The maximum value
$m^{\mu}_i = 1$ arises when $d^{\mu}_j = d^{\mu}_i$ $\forall j \in
{\mathcal V}_i$. The presence of connected domains where the fixed
point $\zeta$ overlaps with the $\xi^\mu$ pattern should be detected
as short range correlations between the local magnetizations. The
correlation between the local magnetizations of the difference
vector with the $\mu$ pattern are then defined as:
\begin{equation}
C^{\mu} = \frac{1}{N} \sum_{i=1}^{N} {\,\,\, \frac{1}{k_i}
\sum_{j \in {\mathcal V}_i} {m^{\mu}_i m^{\mu}_j}}.
\end{equation}
As we intend to capture the existence of correlations in the
difference with patterns that appear in the mixture that makes up the fixed
point $\zeta$, we define the maximum correlation
\begin{equation}
C = \max_{\mu} \left\{ C^{\mu} \right\}.
\label{eqmaxc}
\end{equation}

\begin{figure}
\centering
\resizebox{\columnwidth}{!}{\rotatebox{-90} {\includegraphics{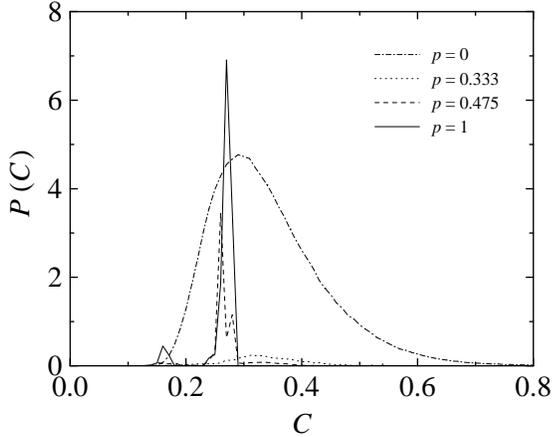}}}
\caption{Distribution of the local correlation that characterizes the level of
alignment with a stored pattern [Eq.~(\ref{eqmaxc})]. System parameters:
$N=2 \times 10^3$, $K=100$, $10^6$ realizations per curve. A peak at $C=1$, shared by the
three curves with the higher values of $p$, is not shown for reasons of scale
(see
discussion in the text).}
\label{cordis}
\end{figure}

Figure~\ref{cordis} presents the probability distribution $P(C)$
for different levels of network disorder. Each distribution is
constructed over $10^6$ realizations of $N=2 \times 10^3$ networks, with
connectivity $K=100$. For $p=0$ we observe a broad peak centered
around $C \approx 0.3$. This is a quantitative measure of the
occurrence of correlations on ordered networks, as we pointed out.
For the other values of $p$ considered in the figure, the
distribution has a sharp peak at $C=1$ which we have not shown for
reasons of scale, corresponding to the fixed points that coincide
with a pattern, and consequently give the highest possible value
of the correlation. Besides this peak, the most disordered systems
show a narrow peak at $C\approx 0.25$, and the curve for $p=1$
also a smaller one at $C\approx 0.15$. These two peaks correspond
to symmetric and asymmetric mixtures, respectively. For $p=0.333$,
very close to the critical point, the distribution presents a very
small bump at $C \approx 0.3$. It is easy to see, form the
extended region of $P(C)$ in the curve for $p=0$, that the
mixtures are characterized by higher local correlation in the
ordered system than in the disordered ones.

\section{Discussion}

We have studied a model of associative memory based on neural
networks with a complex topology. This kind of connectivity can
be considered as more similar to the biological networks than the
completely connected or randomly diluted networks. Many of the
general features of these systems are preserved: the network is
able to retrieve a memorized pattern, up to a saturation.
Besides, we have found a critical dependence of the efficacy of
retrieval on the disorder parameter of the network: a collapse of
the memory capability takes place at a finite value of the
disorder parameter. The optimal performance of the system occurs
at an intermediate value of the disorder, just above the critical
value. This enhanced performance occurs far away from the region
of $p=1$, which is equivalent to the well known models of
completely connected or randomly connected neural networks. We
have characterized the different phases by the properties of the
mixture states, that prevent the system to reach one of the
memorized states.

We have understood the failure of the more ordered networks to
retrieve a stored pattern due to the partition of the system into
arbitrary neighborhoods aligned with more than one pattern. This
is something that the disordered networks cannot do, and in fact
the distributions of the overlaps and of the correlations
quantify this effect. It does not escape us that we cannot, at
this stage, provide an explanation of the enhanced performance of
the intermediate region.

We have checked the robustness of our results with respect to a
small amount of noise in the dynamics. This has been implemented
by flipping, with probability $\epsilon$, one neuron at random
after each deterministic step. For values of $\epsilon$ up to
$0.01$, the results are indistinguishable from the noiseless
system. For greater values of $\epsilon$ the system becomes more
and more ineffective to retrieve a pattern, but the general form
of the curves $\varphi (p)$ is preserved for the whole range of
$p$. A systematic analysis of the problem of a truly noisy
network, characterized by a temperature, remains to be done.

\begin{acknowledgement}
The authors acknowledge fruitful discussions with D. H. Zanette.
G.A. thanks the Abdus Salam ICTP for its hospitality, and
Fundaci{\'o}n Antorchas for financial support.
\end{acknowledgement}

\end{document}